\documentclass[twocolumn]{article}

\usepackage[utf8]{inputenc} 
\usepackage[T1]{fontenc}    
\usepackage{hyperref}       
\usepackage{url}            
\usepackage{booktabs}       
\usepackage{amsfonts}       
\usepackage{nicefrac}       
\usepackage{microtype}      
\usepackage{graphicx}       
\usepackage{amsmath,amsfonts}
\usepackage{algorithmic}
\usepackage{algorithm}
\usepackage{array}
\usepackage{subfig}
\usepackage{textcomp}
\usepackage{stfloats}
\usepackage{verbatim}
\usepackage{cite}
\usepackage{upgreek}
\usepackage[normalem]{ulem}
\usepackage{multirow}
\usepackage{xcolor, colortbl}
\usepackage{siunitx}
\usepackage{bm}
\usepackage[capitalize]{cleveref}
\usepackage{fancyhdr}

\definecolor{LightGray}{gray}{0.9}
\definecolor{CornflowerBlue}{RGB}{100, 149, 237}
\definecolor{ForestGreen}{RGB}{34, 139, 34}

\date{}

\title{Quantum Communication Multiplexing in LP-modes Enabled by Photonic Lanterns}

\author{
  I. Beraza\textsuperscript{1}, 
  M. Zahidy\textsuperscript{1}, 
  R. Mueller\textsuperscript{1}, 
  N. M. Mathew\textsuperscript{1}, 
  L. Grüner-Nielsen\textsuperscript{2}, \\
  L. S. Rishøj\textsuperscript{1}, 
  L. K. Oxenløwe\textsuperscript{1}, 
  M. Galili\textsuperscript{1} \\
  \small \textsuperscript{1} Department of Electrical and Photonics Engineering, Technical University of Denmark, 2800 Kgs. Lyngby, Denmark \\
  \small \textsuperscript{2} Danish Optical Fiber Innovation, Åvendingen 22A, 2700 Brønshøj, Denmark \\
  \small \texttt{imabe@dtu.dk}
}

\begin{document}
\maketitle

\begin{abstract}
The non-cloning theorem of quantum states provides security, but also limits the Secret Key Rate (SKR) for Quantum Key Distribution (QKD) implementations. Multiplexing is a widely used technique to enhance data rates in classical communication systems and can also increase the SKR in QKD systems. Using linearly polarized (LP) modes is an attractive solution as it is compatible with simple fiber designs. This work demonstrates a fiber-based QKD system employing LP mode multiplexing with a photonic lantern to convert the fundamental mode ($LP_{01}$) in separate fibers into higher-order modes ($LP_{11}$) in a single few-mode fiber. The performance of the system is sensitive to polarization dependence, mode alignment, and environmental crosstalk, which requires precise polarization control to minimize Quantum Bit Error Rate (QBER). We report a SKR of \SI{2.34}{Mbps} over a \SI{24}{km} (\SI{5}{dB} loss) fiber link.
\end{abstract}


\section{Introduction}

Quantum Key Distribution (QKD) is a promising approach to secure communication in the era of quantum computing, complementing post-quantum cryptography (PQC) by enabling theoretically unbreakable one-time pad encryption as well as rapid key updates in symmetric cryptography schemes. QKD protocols rely on quantum mechanics principles that prevent copying and amplification of unknown quantum states, typically implemented using single photons. However, a significant challenge in current QKD systems is the low secret key rate (SKR) due to channel loss, which diminishes the rate of key generation compared to classical communication.

To overcome this limitation, a range of solutions have been proposed. These include increasing generation rates \cite{IslamHiDHighRate}, frequency and time multiplexing \cite{Eriksson2019,  Park2022}, spatial mode and path multiplexing \cite{8527341, ZahidyOAMMultiplex}, measurement-device-independent protocols \cite{PhysRevLett.130.210801}, high-dimensional QKD \cite{HybridHiD, Zahidy2024}, satellite-based QKD \cite{Liao2017}, and more advanced technologies such as quantum memories and entanglement swapping \cite{Gera2024}. In fiber-based systems, multiplexing using linearly polarized (LP) modes may be a particularly attractive solution due to compatibility with simple conventional fibers and the availability of fiber-based multiplexers.

In this work, we demonstrate the feasibility of using LP modes along with photonic lanterns (PLs) to multiplex time-bin \cite{Bacco2019FieldTO, ribezzo2023deploying} qubits for QKD. PLs enable fiber-based, low-complexity mode multiplexing, facilitating the transmission and separation of $LP_{11a}$ and $LP_{11b}$ modes in a two-mode graded-index (TMGI) fiber.

This is an extension of our previous work \cite{BerazaECOC24}, which confirmed the feasibility of using LP mode multiplexing for QKD. Here, we add validation of the system including an active decoy-state implementation which allows for accurate channel parameter estimation and enhanced SKR stabilization. This is particularly important when investigating QKD in a channel with time-varying performance, as in this case, where mode rotation occur during propagation in the TMGI fiber. This aspect of the investigation is addressed in detail in section~\ref{results}.

\section{Photonic Lantern}

A photonic lantern (PL) is a mode multiplexer, enabling efficient coupling between single-mode and multimode fibers. It is fabricated by tapering down a bundle of single-mode fibers (SMFs) until the individual cores merge into a common multimode structure, bounded by the air-cladding interface. The tapered bundle is then cleaved and spliced to a multimode fiber. Ideally, light launched into one of the SMFs couples exclusively to a specific higher-order mode in the multimoded fiber. Conversely, when operated in the reverse direction, light from a particular mode in the multimode fiber couples selectively to the corresponding SMF.

The PLs used in this study are designed to excite the first two LP mode groups. The multimode fiber is a two-mode step index (TMSI) fiber, while two different types of SMFs are utilized: a Thorlabs SM2000 and two OFS Clearlite-16 (CL). Light launched into the OFS Clearlite-16 fibers excite an orthogonal combination of theLP$_{11a}$ and LP$_{11b}$ modes, whereas light coupled into the Thorlabs SM2000 fiber will excite light in the fundamental mode of the TMSI \cite{PL}. 

\begin{figure*}[ht]
\centering
\includegraphics[width=0.95\textwidth]{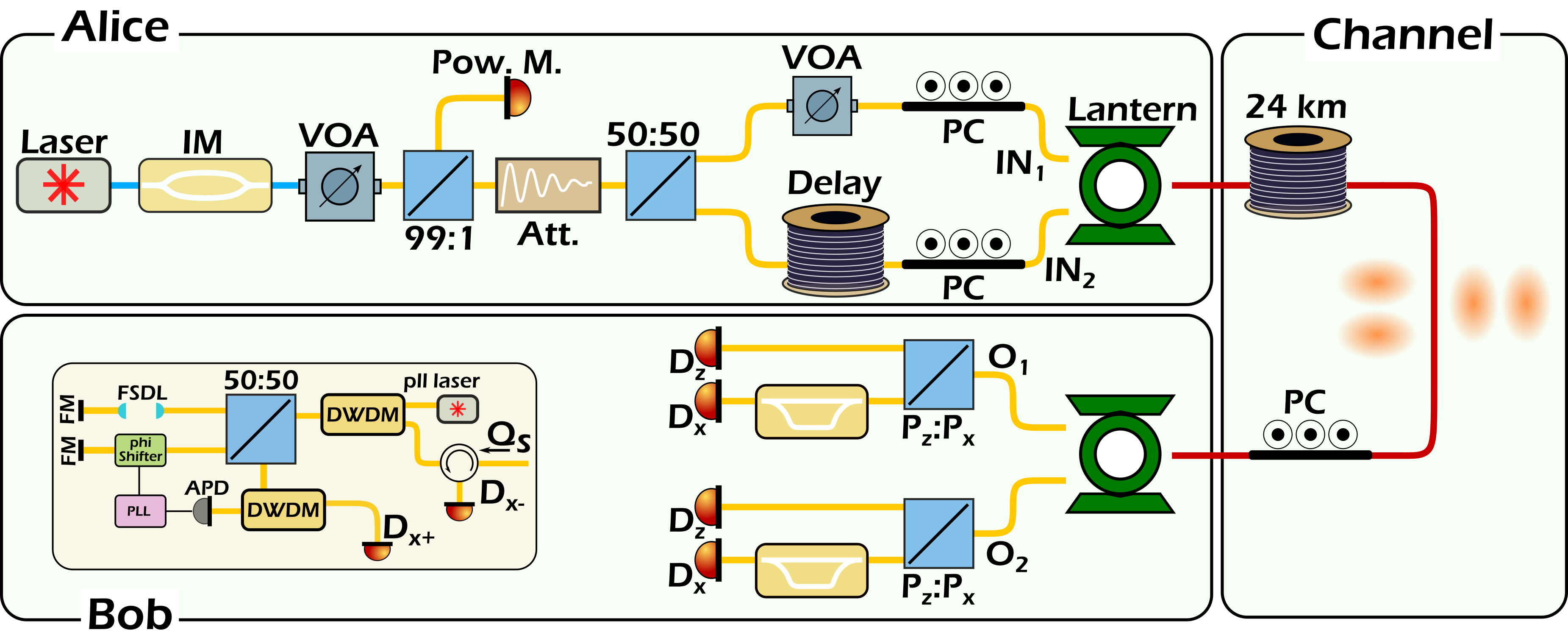}
\caption{\textbf{Alice} (transmitter); IM: intensity modulator, VOA: variable optical attenuator, 99/1: \SI{20}{dB} coupler used to monitor the average power, Att.: Fixed attenuator set to \SI{10}{dB}, PC: polarization controller, Lantern: photonic lantern. \textbf{Bob} (Receiver); DWDM: Dense wavelength-division multiplexing, APD: Avalanche photo-diode, Phi-Shifter: fiber stretcher to adjust the phase, FSDL: Free space delay line, FM: Faraday mirror, PLL: Phase locked Loop.}
\label{FIG::Setup}
\end{figure*}

In the QKD system, PLs are used for both multiplexing and de-multiplexing. The efficiency of the multiplexing  PL is characterized using the spatial and spectral imaging (S$^2$) method \cite{Nicholson08S2}. From this measurement the crosstalk is determined, which is defined as the ratio of power in the undesired mode to the power in the target mode. The crosstalk for the multiplexing PL is measured at 1550 nm for the three SMF inputs for ten random input polarization states. The results are shown in \cref{MuXPL2}.

\begin{table}[h!]
\centering
\caption{Measured crosstalk for the multiplexing Photonic Lantern.}
\renewcommand{\arraystretch}{1.3} 
\arrayrulecolor{ForestGreen}     

\resizebox{\columnwidth}{!}{
\begin{tabular}{lccc}
\toprule
\textbf{Input Port} & \textbf{Min [dB]} & \textbf{Max [dB]} & \textbf{Average [dB]} \\ \midrule
SM2000              & -15.5             & -14.1             & -14.7                \\
OFS CL1             & -20               & -14.9             & -17.5                \\
OFS CL2             & -22.2             & -15.3             & -18                  \\
\bottomrule
\end{tabular}
} 
\arrayrulecolor{black} 
\label{MuXPL2}
\end{table}

The polarization-dependent insertion loss (PDL) for the multiplexing PL is characterized using the polarization scanning method \cite{PDLscanning}. For this measurement light is launched through one of the three input SMF ports and the output power of the TMSI fiber is monitored while the input polarization is randomly varied for a duration long enough to cover the whole Poincaré sphere. The measured PDL for the SM2000, OFS CL1, and OFS CL2 ports are 2.1-2.4 dB, 4.3-4.4 dB, and 5.5-5.7 dB, respectively. 

The OFS CL fibers excite a combination of the LP$_{11a}$ and LP$_{11b}$ modes.  However, the LP$_{11}$ components excited by the two OFS CL fibers are orthogonal, enabling multiplexing in the degenerate LP mode basis of the TMSI fiber. The demultiplexing PL is characterized by measuring the insertion loss for the LP$_{01}$ and LP$_{11}$ modes in the demultiplexing direction. For this measurement, the LP$_{01}$ mode is launched in the TMSI fiber of the PL through a well-aligned splice to an SMF, furthermore, any undesired higher-order modes content is removed using a mode stripper \cite{LarsOFT}. The output power is then measured at each of the three SMF ports of the PL while scanning the input polarization states. The powers of the two OFS CL fibers are combined to calculate the total power in the LP$_{11}$ mode. To characterize the LP$_{11}$ mode, light is launched into the specific mode of the TMSI using a long period grating \cite{LarsLPG}. The highest demultiplexing crosstalk observed for the $LP_{01}$ and $LP_{11}$ modes was measured at $-11.3$ dB and $-14.6$ dB, respectively, indicating the worst-case scenarios for interference between the modes. 

Note that as the $LP_{11}$ mode propagates through the fiber the mode orientation will continuously rotate. However, by mounting the multimoded fiber in a polarization controller it is possible to align the mode orientation to the preferred orientation of the PL, hereby ensuring optimal demultiplexing. This follows the approach outlined in \cite{PolCon}.

\section{Time-bin QKD with Photonic Lanterns} 
\label{setup}

We implemented a QKD system and multiplexed the transmission of quantum signals in spatial modes via the photonic lantern. The protocol is the efficient three-state BB84 time-bin encoding with one decoy state to counter the photon-number splitting (PNS) attack \cite{Rusca_2018}. \cref{FIG::Setup} depicts the experimental setup. Time-bin qubits are generated at a rate of \SI{1.25}{GHz} by carving a continuous wave (CW) laser, which is set at a wavelength of \SI{1550.92}{nm} - channel 33 of the International Telecommunication Union-Telecommunication Standardization Sector (ITU-T). The carving is performed using Lithium Niobate Mach Zehnder modulators to guarantee a high extension ratio. A field programmable gate array (FPGA) generates the electrical signal used for carving according to a pre-loaded sequence of length $l=5000$ random states. 
This sequence comprises $\mathcal{Z}$ and $\mathcal{X}$ states in a $90/10$ ratio. The $\mathcal{Z}$ states, $\{|e\rangle, |l\rangle\}$, are used for key generation and the $\mathcal{X}$ state, $\{ |e\rangle \otimes |l\rangle \}$ is employed for security purposes \cite{Rusca_2018}. In this notation, $|e\rangle$($|l\rangle$) is a weak coherent pulse occupying the early(late) time-bin, respectively.

The signal and decoy states are generated by applying a 3-level signal to the cascaded intensity modulators resulting in two different levels. The signal is engineered to generate a ratio of 3.1 between $\mu_1$ and $\mu_2$.
A series of variable optical attenuators (VOA), a \SI{20}{dB} coupler, and a fixed attenuator reduce the mean photon number per qubit to below one, while enabling power monitoring to set the mean photon number to desired values before the input of the channel.
For security, consecutive states should be phase-randomized and selected fully at random. This calls for the implementation of a quantum random number generator \cite{QRNG}. 
To perform multiplexing, the signals are divided using a 50:50 beam splitter and decorrelated using a delay line to remove any coherence between the two ports. These signals are then directed into the input ports of the PL, converting them into a superposition of $LP_{11a}$ and $LP_{11b}$ spatial modes. Two PCs are used to optimize spatial mode fidelity, i.e. minimize crosstalk. A VOA balances the intensity of the modes to ensure an equal mean photon number for both channels.

The converted signals propagate through \SI{24}{km} of the TMGI fiber. At the receiver, the second PL acts as a demultiplexer, separating the two modes and converting them back to the fundamental mode. Demultiplexing efficiency, with minimal crosstalk, depends on the polarization state of the light and the orientation of the modes. The orthogonality between the incoming modes ($LP_{11a/b}$) enhances the effectiveness of the demultiplexing process. Therefore, a polarization controller positioned immediately before the PL is critical for ensuring proper mode separation \cite{PolCon}. The signals are subsequently passed through a beam splitter with $P_z: P_x$ splitting ratio where they are subjected to $\mathcal{Z}$ and $\mathcal{X}$ measurement, respectively.
\begin{table}[]
\centering
\caption{Experimental parameters, QBERs detection rates, and secret key rates (block size: $10^9$). The estimated channel loss is \SI{5}{dB}.}
\arrayrulecolor{ForestGreen}
\label{tab:ParamAggregate}
\resizebox{\columnwidth}{!}{%
\begin{tabular}{lll|ll}
\multicolumn{1}{l}{} & \multicolumn{2}{l|}{\textbf{No Decoy}} & \multicolumn{2}{l}{\textbf{Active Decoy}} \\ \cline{2-5} 
                                & $\mu_1$ & $\mu_2$ & $\mu_1$ & $\mu_2$ \\ \hline
Mean Photon Number              & 0.6     & 0.2     & 0.31     & 0.1     \\
$P_{\mu}$                       & 0.8     & 0.2     & 0.8     & 0.2     \\
$P_{\mathcal{Z}/\mathcal{X}}^A$ & 80/20   & 80/20   & 90/10   & 90/10   \\
$P_{\mathcal{Z}/\mathcal{X}}^B$ & 50/50   & 50/50   & 75/25   & 75/25   \\
$Q_\mathcal{Z}$ Mode 1 [\%]     & 1.3     & 0.86    & 4.43    & 4.43    \\
$Q_\mathcal{Z}$ Mode 2 [\%]     & 4.35    & 3.85    & 5.56    & 5.56    \\
$Q_\mathcal{X}$ Mode 1 [\%]     & 2.7     & 3.22    & 2.14    & 2.14    \\
Det. Rate $\mathcal{Z}$ Mode 1 [MHz] & 1.73   & 0.25      &  3.78     & 0.32      \\
Det. Rate $\mathcal{Z}$ Mode 2 [MHz] & 2.47   & 0.18      &  1.97     & 0.17      \\
SKR [Mbps] & \multicolumn{2}{c|}{\textbf{2.23}} & \multicolumn{2}{c}{\textbf{2.34}}         
\end{tabular}%
}
\arrayrulecolor{black}
\end{table}
In the $\mathcal{Z}$-basis measurement, a single-photon detector registers the time of arrival of the quantum states. The $\mathcal{X}$-basis measurement, on the other hand, is performed with an unbalanced Michelson interferometer with 400 ps delay in one arm. The structure of the fiber-based interferometer is depicted in \cref{FIG::Setup}. It is comprised of a 50:50 beam splitter, two Faraday rotator mirrors, a free-space delay line to temporally match the length of the two arms, and a phase shifter to stabilize the two arms of the interferometer. The interferometer is stabilized with the dual-band technique where a secondary laser, set to wavelength \SI{1554.13}{nm} -channel 29 ITU-T, is co-propagated with the quantum signal and is used to monitor the phase fluctuation. A phase-locking loop then stabilizes the phase by applying the right phase shift via a fiber stretcher. 

Superconducting nanowire single-photon detectors (SNSPD) $D_Z$ and $D_X$, are used to register the quantum signals with $\approx 50$ dark counts per second,  \SI{33}{ns} dead time and \SI{83}{\percent} detection efficiency. Detection events are recorded by a time-to-digital converter (TDC) with \SI{1}{ps} resolution for post-processing. Synchronization between the transmitter and receiver is maintained by a down-sampled \SI{250}{kHz} clock signal provided by the FPGA.

\section{Results and discussion} \label{results}

For many QKD system investigations, emulating the decoy states, which is necessary to protect against PNS attacks, can be implemented by subsequently testing the system with different mean photon numbers. As long as the channel conditions are stable this is sufficient to predict the achievable SKR. In our system, we observed that time-varying crosstalk, which is likely to occur in few-mode fiber transmission channels, causes inconsistencies in QBER and detection rates, which for secret key production must be considered as ongoing PNS attacks. Under these conditions, active decoy state generation becomes a necessity to accurately characterize the system. 

A change in the rate of received qubits while performing data acquisition for different mean photon numbers, $\mu$, separately, can lead to an inconsistent reading of the detection rate. This, in turn, drastically reduces the final SKR as it appears as an ongoing PNS attack. To avoid this, an active implementation of the decoy state is necessary. For long-term operations, an active mode-control to reduce the crosstalk is paramount, however, this is beyond the scope of this paper. 
\begin{figure}[H]
    \centering
    \includegraphics[width=0.9\linewidth]{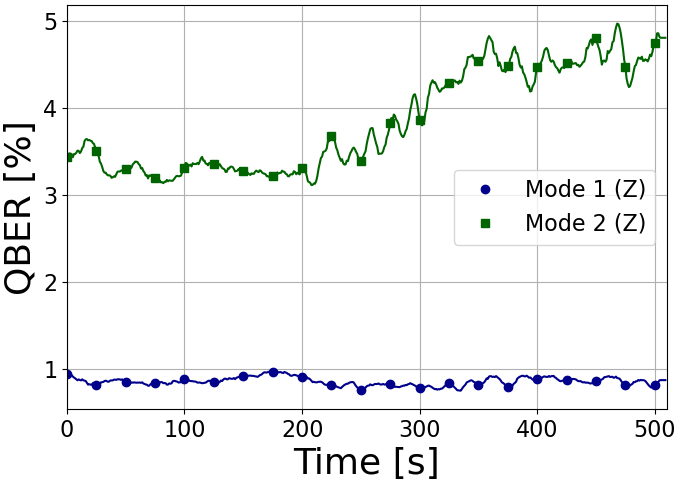}
    \caption{Variation of QBER for the experiment without active decoy state, for $\mu = 0.2$.}
    \label{fig::QBERvsTime}
\end{figure}

In the following, we present our results before and after implementing an active decoy state generation. Table \ref{tab:ParamAggregate} summarizes the parameters chosen in each experiment, including the observed quantum bit error rates, detection rates, and, estimated secret key rates. The values $\mu_i$ and $P_{\mu_i}$ are of mean photon number measured at the output of the photonic lantern and prior to the TMGI and the probability of transmission/selection for signal and decoy states, respectively. $P_{\mathcal{Z}}^A$($P_{\mathcal{Z}}^B$) is the probability that Alice(Bob) choose the $\mathcal{Z}$ basis for state generation(measurement) with $P_{\mathcal{X}} = 1 - P_{\mathcal{Z}}$. $Q_\mathcal{Z}$ and $Q_\mathcal{X}$ denote the average value of QBER measured in each basis, for each mode. The values of $P_{\mu_i}$, and $P_{\mathcal{Z}}^A$ are numerically optimized for the QBER measured in the respective experiment.

Due to limited detector availability, measurements were restricted to a $\mathcal{Z}$ and $\mathcal{X}$ bases for Mode~1, while Mode~2 was measured only in the $\mathcal{Z}$ basis.

To minimize crosstalk and achieve low QBER during QKD multiplexing, the polarization controllers were adjusted. The PCs at the input of the photonic lantern were initially tuned, followed by adjustments of the PC in the channel. This procedure was repeated iteratively, alternating between the input and the PC on the TMGI fiber, until a configuration with minimal QBER was achieved. Once the system reached a stable low QBER state, data acquisition commenced, with continuous monitoring to ensure stability during the measurements.

Figure \ref{fig::QBERvsTime} shows the QBER measured over a period of 500 seconds. The slow variation in QBER and the low values, mainly for Mode 1, suggest that LP-mode multiplexing in few-mode fibers is a promising method to increase the channel capacity if the modal crosstalk is actively monitored and compensated for. Sequential data acquisition for $\mu_1$ and $\mu_2$ experienced variations in channel conditions between measurements. Specifically, channel conditions during the $\mu_2$ measurement in Mode 2 differed slightly from those during the $\mu_1$ measurement. These inconsistencies reduced the SKR since such variations could theoretically be exploited by an eavesdropper, thereby lowering the number of detections that are confidently attributed to single-photon pulses.

\begin{figure}
    \centering
    \includegraphics[width=0.9\linewidth]{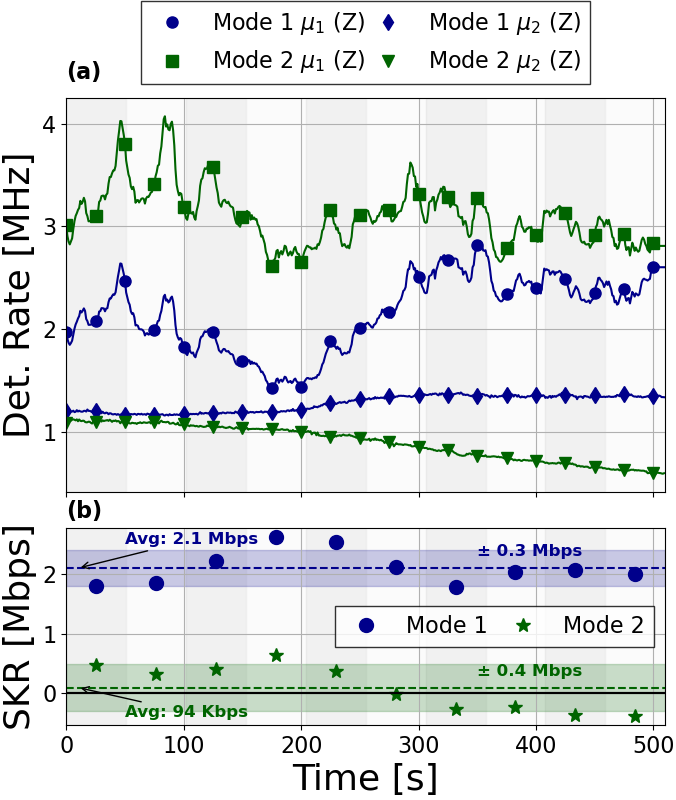}
    \caption{\textbf{a)} Variation of detection rate. \textbf{b)} Average SKR for each period of \SI{50}{s} of data, or the experiment without active decoy state. Mode 1: $\pm$ \SI{0.3}{Mbps}  of standard deviation. Mode 2: $\pm$ \SI{0.4}{Mbps} of standard deviation. Block size: $10^8$.}
    \label{fig::DetRateSKRvsTime}
\end{figure}

Figure \ref{fig::DetRateSKRvsTime}(a) depicts the detection rate for both modes and for two different mean photon number measured over time. We estimated, Eq. (\ref{Eq::L_finite}), the extractable SKR for 10 sub-periods (indicated by the alternating white and gray shaded regions) with corresponding block sizes, see Fig. \ref{fig::DetRateSKRvsTime}(b). Over time, the estimated SKR of Mode 2 falls below zero due to crosstalk, leading it to be interpreted as a PNS attack, despite the average QBER remaining at an acceptable level of approximately 4\%, see \cref{tab:ParamAggregate}. Note that the $LP_{01}$ mode was not monitored due to lack of detector availability, however, we attribute the change of detection rate to overall crosstalk between all three modes.

\begin{figure}
    \centering
    \includegraphics[width=\linewidth]{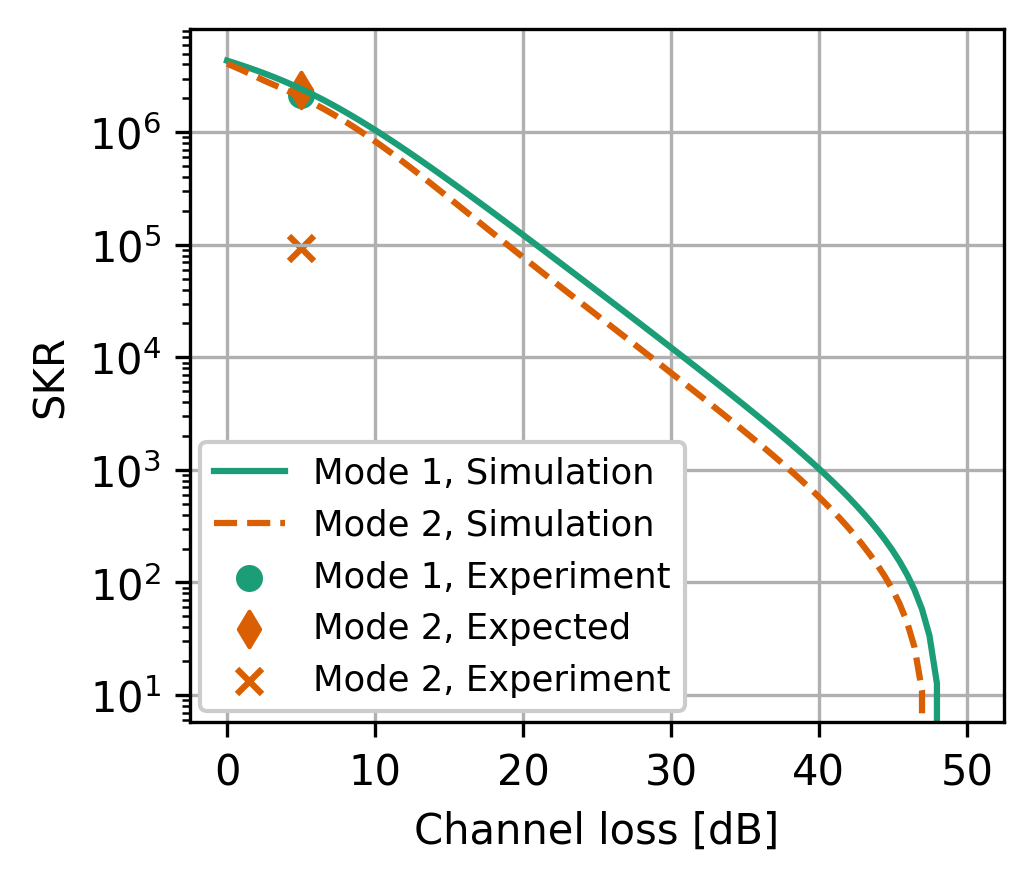}
    \caption{SKR versus channel losses. The markers represent the SKR at \SI{5}{dB} channel loss for \SI{24}{km} of two-mode fiber.}
    \label{fig::SKR}
\end{figure}

The measured SKR was \SI{2.13}{Mbps} for Mode 1 and \SI{97}{kbps} for Mode 2. Under consistent conditions where both mean photon numbers were measured simultaneously, the SKR for Mode 2 was estimated to reach \SI{2.35}{Mbps}, yielding a total SKR of \SI{4.58}{Mbps}. Figure \ref{fig::SKR} shows the measured SKR as a function of channel loss, highlighting the expected decline at higher losses.
\begin{figure}
    \centering
    \includegraphics[width=0.9\linewidth]{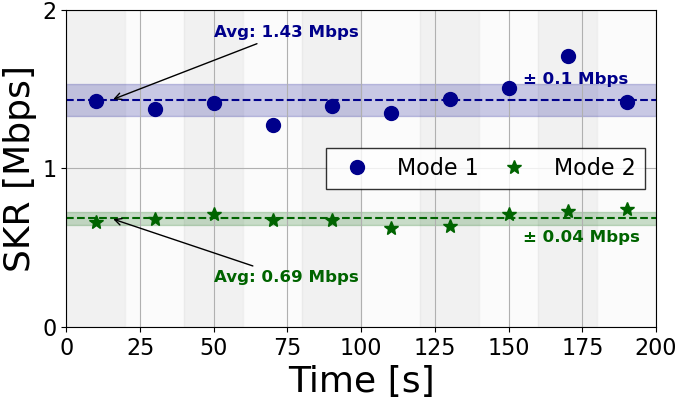}
    \caption{Average SKR for each period of \SI{50}{s} of data, for the experiment with active decoy state. Mode 1: $\pm$ \SI{0.1}{Mbps} of standard deviation. Mode 2: $\pm$ \SI{0.04}{Mbps} of standard deviation. Block size: $10^8$.}
    \label{fig::DetRateSKRvsTime_decoy}
\end{figure}

The active decoy generation was implemented and set to generate two intensities of ratio 3.1. In this case, time-varying crosstalk affects both signal and decoys as an equal change in the channel transmission eliminating the PNS assumption in the SKR estimation. For experimental parameters, refer to Table \ref{tab:ParamAggregate}. As in the previous configuration, the input PCs and PC in the channel were carefully adjusted to minimize crosstalk, and this configuration was maintained throughout the data acquisition. 

The SKR is estimated at \SI{1.6}{Mbps} for Mode 1 and \SI{0.74}{Mbps} for Mode 2, resulting in a total SKR of \SI{2.34}{Mbps}. We observe a more stable SKR, see \cref{fig::DetRateSKRvsTime_decoy}, due to the elimination of the effect of crosstalk on the final SKR. Comparing \cref{fig::DetRateSKRvsTime}b) and \cref{fig::DetRateSKRvsTime_decoy}, specifically for Mode 2 a decrease in the standard deviation is observed, from \SI{0.4}{Mbps} for the non-decoy experiment to \SI{0.04}{Mbps} for the decoy one.

In both cases, the secret key length $l$ in the finite-key regime was estimated using the expression \cite{Rusca_2018}:

\begin{equation}
\begin{split}
l_{\text{Finite}} &\leq D_0^Z + D_1^Z[1 - h(\phi_Z)] - \lambda_{EC} \\
&\quad - 6 \log_2\left(\frac{21}{\epsilon_{\text{sec}}}\right) - 2 \log_2\left(\frac{2}{\epsilon_{\text{corr}}}\right).
\end{split}
\label{Eq::L_finite}
\end{equation}

A meaningful comparison lies in the relative performance between the two modes within each case. In the absence of active decoy generation, there is a significant disparity in SKR between the two modes. This difference is much less pronounced in the active decoy generation case. These results indicate that for systems susceptible to environmental fluctuations and crosstalk, the implementation of active decoy generation enhances the key generation rate.

\section{Conclusions}

We have demonstrated time-bin QKD multiplexed in LP modes enabled by photonic lanterns. The system inherently exhibits QBER stability over 10 minutes, with levels below \SI{4}{\percent}, allowing us to perform this first demonstration without active system stabilization. The demonstrated stable SKR is \SI{2.3}{Mbps} when implementing active decoy state generation and is achieved over a distance of \SI{24}{km} in FMF. In this work we confirm the feasibility of QKD using fiber-based PLs and LP modes in FMF. As PL technology advances, this platform appears quite promising, offering an all-fiber solution to QKD multiplexing in a single fiber core. 

Our investigation revealed that crosstalk fluctuations during QKD operation may cause detection rate fluctuations, which in turn produces conditions similar to an ongoing PNR attack when performing the secret key generation. The origin of these fluctuations is linked to the inevitable mode- and polarization changes occurring in the fiber system due to changing ambient conditions. We analyze the phenomenon and confirm that implementation of active decoy state generation allows for stable secret key generation even in the presence of these fluctuations in the fiber channel supporting QKD. 
This observation highlights the importance of active decoy generation when studying such systems, in contrast to the often valid approach of testing the system with different mean photon numbers sequentially.

This feasibility study of the use of $LP_{11}$ modes for QKD multiplexing does not address long-term operation. However, the slow variation of the QBER during measurements suggests that active stabilization of polarization- and mode rotations could stabilize and maintain low QBER levels, thereby enhancing the SKR and enabling secure key exchange over extended periods.

Although our experiments focused solely on the $LP_{11}$ modes, other modes could be included either to achieve higher levels of multiplexing, higher-dimensional QKD protocols, co-propagation of classical and quantum signals, and the provision of classical service channels in the same fiber used for QKD. As such, we believe that LP modes offer many interesting possibilities for QKD systems.

\bibliographystyle{unsrt}
\bibliography{References}

\begin{thebibliography}{10}

\bibitem{IslamHiDHighRate}
N.~T. Islam, C.~C.~W. Lim, C.~Cahall, J.~Kim, and D.~J. Gauthier.
\newblock Provably secure and high-rate quantum key distribution with time-bin qudits.
\newblock {\em Science Advances}, 3(11):e1701491, 2017.

\bibitem{Eriksson2019}
T.~A. Eriksson, T.~Hirano, B.~J. Puttnam, and et~al.
\newblock Wavelength division multiplexing of continuous variable quantum key distribution and 18.3 tbit/s data channels.
\newblock {\em Communications Physics}, 2(1):9, 2019.

\bibitem{Park2022}
C.~H. Park, M.~K. Woo, B.~K. Park, and et~al.
\newblock 2×n twin-field quantum key distribution network configuration based on polarization, wavelength, and time division multiplexing.
\newblock {\em npj Quantum Information}, 8(1):48, 2022.

\bibitem{8527341}
B.~D. Lio, D.~Bacco, D.~Cozzolino, and et~al.
\newblock Record-high secret key rate for joint classical and quantum transmission over a 37-core fiber.
\newblock In {\em 2018 IEEE Photonics Conference (IPC)}, pages 1--2, 2018.

\bibitem{ZahidyOAMMultiplex}
M.~Zahidy, Y.~Liu, D.~Cozzolino, and et~al.
\newblock Nanophotonics.
\newblock {\em vol. 11, no. 4}, pages 821--827, 2022.

\bibitem{PhysRevLett.130.210801}
Y.~Liu, W.~J. Zhang, C.~Jiang, and et~al.
\newblock Experimental twin-field quantum key distribution over 1000 km fiber distance.
\newblock {\em Phys. Rev. Lett.}, 130:210801, 2023.

\bibitem{HybridHiD}
Y.~Jo, H.~S. Park, S.-W. Lee, and W.~Son.
\newblock Efficient high-dimensional quantum key distribution with hybrid encoding.
\newblock {\em Entropy}, 21(1), 2019.

\bibitem{Zahidy2024}
M.~Zahidy, D.~Ribezzo, C.~De Lazzari, and et~al.
\newblock Practical high-dimensional quantum key distribution protocol over deployed multicore fiber.
\newblock {\em Nature Communications}, 15(1):1651, 2024.

\bibitem{Liao2017}
S.~K. Liao, W.-Q. Cai, W.~Y. Liu, and et~al.
\newblock Satellite-to-ground quantum key distribution.
\newblock {\em Nature}, 549(7670):43--47, 2017.

\bibitem{Gera2024}
W.~Li, L.~Zhang, H.~Tan, and et~al.
\newblock High-rate quantum key distribution exceeding 110 mb s–1.
\newblock {\em Nature Photonics}, 17(5):416--421, 2023.

\bibitem{Bacco2019FieldTO}
D.~Bacco, I.~Vagniluca, B.~D. Lio, and et~al.
\newblock Field trial of a three-state quantum key distribution scheme in the florence metropolitan area.
\newblock {\em EPJ Quantum Technology}, 6:1--8, 2019.

\bibitem{ribezzo2023deploying}
D.~Ribezzo, M.~Zahidy, I.~Vagniluca, and et~al.
\newblock Deploying an inter-european quantum network.
\newblock {\em Advanced Quantum Technologies}, 6(2):2200061, 2023.

\bibitem{BerazaECOC24}
I.~Beraza, M~Zahidy, R.Mueller, L.~Grüner-Nielsen, L.~S. Rishøj, L.~K. Oxenløwe, K.~Rottwitt, and M.~Galili.
\newblock Towards quantum communication multiplexing in lp-modes enabled by photonic lanterns.
\newblock In {\em ECOC}, page M3A.5, 2024.

\bibitem{PL}
L.~Gr{\"u}ner-Nielsen, N.~M. Mathew, and K.~Rottwitt.
\newblock Direct measurement of polarization dependency of mode conversion in a long period grating.
\newblock {\em Optical Fiber Communication Conference (OFC)}, page Th2A.15, 2019.

\bibitem{Nicholson08S2}
J.~W. Nicholson, A.~D. Yablon, S.~Ramachandran, and S.~Ghalmi.
\newblock Spatially and spectrally resolved imaging of modal content in large-mode-area fibers.
\newblock {\em Opt. Express}, 16(10):7233--7243, 2008.

\bibitem{PDLscanning}
Y.~Zhu, E.~Simova, P.~Berini, and C.~P. Grover.
\newblock A comparison of wavelength dependent polarization dependent loss measurements in fiber gratings.
\newblock {\em IEEE Transactions on Instrumentation and Measurement}, 49(6):1231--1239, 2000.

\bibitem{LarsOFT}
L.~Gr{\"u}ner-Nielsen, N.~M. Mathew, and K.~Rottwitt.
\newblock Invited paper: Characterization of few mode fibers and devices.
\newblock {\em Optical Fiber Technology}, 52:101972, 2019.

\bibitem{LarsLPG}
N.~Mariam Mathew, J.~Christensen, L.~Grüner-Nielsen, M.~Galili, and K.~Rottwitt.
\newblock Air-cladded mode-group selective photonic lanterns for mode-division multiplexing.
\newblock {\em Optics Express}, 27(9):13329--13343, 2019.

\bibitem{PolCon}
Y.~Li, L.~Feng, S.~Wu, C.~Yang, W.~Tong, W.~Li, J.~Qiu, X.~Hong, Y.~Zuo, H.~Guo, and J.~Wu.
\newblock Realization of linear-mapping between polarization poincar{\'e} sphere and orbital poincar{\'e} sphere based on stress birefringence in the few-mode fiber.
\newblock {\em Optics Express}, 27(24):35537--35547, 2019.

\bibitem{Rusca_2018}
D.~Rusca, A.~Boaron, F.~Grünenfelder, A.~Martin, and H.~Zbinden.
\newblock Finite-key analysis for the 1-decoy state qkd protocol.
\newblock {\em Applied Physics Letters}, 112(17):171104, 2018.

\bibitem{QRNG}
X.~Ma, X.~Yuan, and Z.~et~al. Cao.
\newblock Quantum random number generation.
\newblock {\em npj Quantum Inf}, 2:16021, 2016.

\end{thebibliography}

\end{document}